\documentclass[aps,prl,twocolumn,groupedaddress]{revtex4}

\usepackage{graphicx} 
\usepackage{dcolumn}
\usepackage{bm}

\begin{document}
\title{Unambiguous evidence for nearly isotropic $s$-wave gap in the
bulk of optimally electron-doped Nd$_{1.85}$Ce$_{0.15}$CuO$_{4-y}$ } 
\author{Guo-meng Zhao$^{1,2,*}$} 
\affiliation{$^{1}$Department of Physics and Astronomy, 
California State University, Los Angeles, CA 90032, USA~\\
$^{2}$Department of Physics, Faculty of Science, Ningbo
University, Ningbo, P. R. China}

\begin{abstract}
    
We address 
an important issue 
as to whether bulk-sensitive data of Raman scattering, 
optical conductivity, magnetic penetration depth, 
directional point-contact
tunneling spectra, and nonmagnetic pair-breaking effect in optimally 
electron-doped Nd$_{1.85}$Ce$_{0.15}$CuO$_{4-y}$  support
a nodeless $s$-wave or $d$-wave superconducting gap. We numerically calculate Raman
intensities, directional 
point-contact tunneling spectra, and
nonmagnetic pair-breaking effect in terms of both $s$-wave 
and $d$-wave gap symmetries. We find that all these bulk-sensitive data are in 
quantitative
agreement with a nearly isotropic $s$-wave gap. The fact that $T_{c}$ is
nearly independent of the residual resistivity rules out any $d$-wave 
gap symmetry.

\end{abstract}

\maketitle

\vspace{0.3cm}


The gap  symmetry of  high-temperature cuprate 
superconductors has been a topic of intense debate for over twenty 
years. For hole-doped cuprates,  bulk-sensitive experiments 
probing low-energy excitations in the superconducting state have consistently 
pointed towards the existence of line 
nodes in the gap function of hole-doped cuprates
\cite{Hardy,Jacobs,Lee,Chiao}, which is consistent with either
$d$-wave gap (having four line nodes) or extended $s$-wave gap (having
eight line nodes). Other bulk-sensitive experiments on hole-doped
cuprates \cite{Bha,Li} can be quantitatively explained by extended $s$-wave gap
\cite{Zhao2001}. In contrast, the issue as to whether there exist
line nodes in the gap function of electron-doped ($n$-type) cuprates remains
controversial. Phase and surface-sensitive
experiments \cite{Tsu} provided evidence for pure
$d$-wave order-parameter (OP) symmetry in optimally doped and overdoped $n$-type cuprates.
Surface-sensitive angle-resolved
photoemission spectroscopy (ARPES) \cite{Arm01,Matsui} showed a
$d$-wave gap with a maximum gap size of about 2.5 meV.  This gap size 
would imply a $T_{c}$ of about 14 K at the surface, which is significantly
lower than the bulk $T_{c}$ of 26 K (Ref.~\cite{Matsui}).  Earlier magnetic penetration 
depth data \cite{Alff} of optimally electron-doped
cuprates with $T_{c}$ = 24 K were shown to be consistent 
with nodeless $s$-wave gap
symmetry. Later on, a $T^{2}$ dependence of 
the penetration depth  at low
temperatures was observed in Pr$_{1.85}$Ce$_{0.15}$CuO$_{4-y}$ ($T_{c}$ =
20 K), which
appears to support $d$-wave gap symmetry in the dirty limit \cite{Pr}. 
But the same data can be quantitatively 
explained \cite{Zhao2001} in terms of a nodeless $s$-wave gap if one takes into account an 
extrinsic effect due to current-induced nucleation of vortex-antivortex pairs at
defects. Extensive penetration depth data \cite{Kim} of
Pr$_{2-x}$Ce$_{x}$CuO$_{4-y}$
confirmed the nodeless gap symmetry at all the doping levels
except for  a deeply underdoped Pr$_{1.885}$Ce$_{0.115}$CuO$_{4-y}$
sample with $T_{c}$ = 12 K. Point-contact tunneling spectra \cite{Kas,Bis,Qaz,Shan05,Shan08}
also showed no
zero-bias conductance peak (ZBCP) at all the doping levels except for
a deeply underdoped Pr$_{1.87}$Ce$_{0.13}$CuO$_{4-y}$ with $T_{c}$ = 12 K.
Therefore, the penetration depth and point-contact tunneling spectra
consistently suggest that the gap symmetry in deeply underdoped
samples should be $d$-wave and change to a nodeless $s$-wave when the
doping level is above a critical value. This scenario can naturally explain
the $d$-wave gap symmetry inferred from 
surface-sensitive experiments if surfaces or
interfaces are deeply underdoped. Experiments on
hole-doped cuprates \cite{Bet,Mann} indeed show that surfaces
and interfaces are significantly underdoped. 
 
Here we focus on optimally 
electron-doped Nd$_{1.85}$Ce$_{0.15}$CuO$_{4-y}$ (NCCO) to
unambiguously address this important issue 
as to whether bulk-sensitive data of Raman scattering, 
optical conductivity, magnetic penetration depth, 
directional point-contact
tunneling spectra, and nonmagnetic pair-breaking effect support
a nodeless $s$-wave gap. We numerically calculate Raman
intensities, directional 
point-contact tunneling spectra, and
nonmagnetic pair-breaking effect in terms of both $s$-wave 
and $d$-wave gap symmetries. We find that all these bulk-sensitive data are in 
quantitative agreement with a nearly isotropic $s$-wave gap. The fact that $T_{c}$ is
nearly independent of the residual resistivity rules out any $d$-wave 
gap symmetry.

\begin{figure}[htb]
    \vspace{-0.5cm}
    \includegraphics[height=12cm]{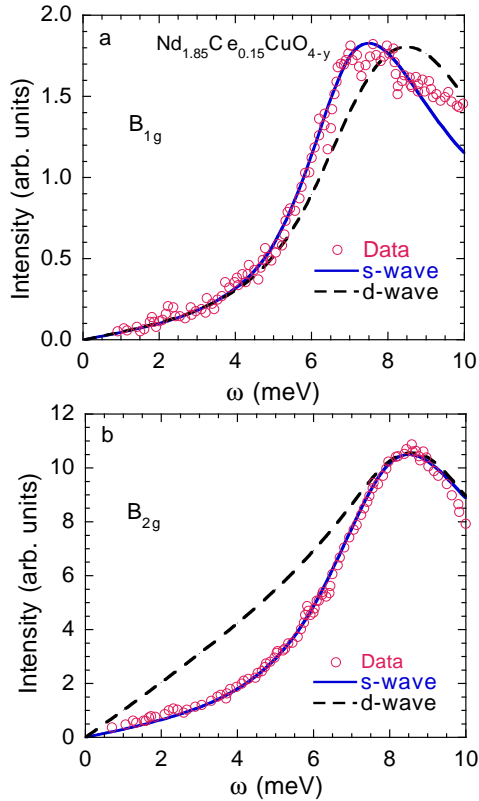}
    \vspace{-0.3cm}
 \caption[~]{ Raman intensities at 8 K for
 Nd$_{1.85}$Ce$_{0.15}$CuO$_{4-y}$ 
($T_{c}$ = 22$\pm$1 K) in the $B_{1g}$ (Fig.~1a) and $B_{2g}$
(Fig.~1b) symmetries. The data are digitized from Ref.~\cite{Blu}. The solid lines 
are numerically calculated
curves in terms of an anisotropic s-wave gap function: $\Delta$ = $3.59
(1-0.11\cos
4\theta)$ meV with $\Gamma$ = 1.4 meV for the $B_{1g}$ channel and 
$\Gamma$ = 1.9 meV for the $B_{2g}$ channel; the dashed lines are the numerically 
calculated curve in terms of a non-monotonic $d$-wave gap function: $\Delta$ 
= $3.25 (1.43\cos 2\theta-0.43\cos 6\theta)$ meV with $\Gamma$ = 1.6 meV for both channels.  }
\end{figure}

Bulk-sensitive Raman scattering has been proved to
be a very powerful tool to study the anisotropy of the
superconducting energy gap. Experiments carried out with different polarization
orientations pick up the contributions to the light scattering
on different parts of the Fermi surface. The $B_{1g}$ spectra provide
information on the light scattering primarily in the neighborhood of the 
$k_{x}$ and $k_{y}$ axes while $B_{2g}$ spectra probe
mainly along the diagonals, where ($k_{x}$, $k_{y}$) = $\vec{k}$ is
the in-plane wave vector of electrons. The electronic Raman intensity is
proportional to the imaginary part of Raman susceptibility
$\chi_{\gamma\gamma} (\vec{q},\omega)$ in the limit of $\vec{q}$
approaching
0. At zero temperature, the imaginary part of $\chi_{\gamma\gamma} (\omega)$
is given by \cite{Branch}
\begin{eqnarray}\label{Raman}
Im\chi_{\gamma\gamma} (\omega) =\sum_{\vec{k}}\frac{\gamma^{2}
(\vec{k})\Delta^{2}(\vec{k})}{E^{2}(\vec{k})}[\frac{\Gamma}{(\omega - 2E(\vec{k}))^{2} + \Gamma^{2}}\nonumber \\
-\frac{\Gamma}{(\omega + 2E(\vec{k}))^{2} + \Gamma^{2}}],
\end{eqnarray}
\noindent where $\Delta (\vec{k})$ is the momentum-dependent superconducting
energy gap, $E(\vec{k})$ = $\sqrt{\Delta^{2}(\vec{k}) +
\epsilon^{2}(\vec{k})}$, $\Gamma$ is the parameter associated with the life-time
broadening of
the quasiparticles, $\epsilon (\vec{k})$ is the band-dispersion relation, and  $\gamma (\vec{k})$ is the Raman vertex 
which is
proportional to $\cos k_{x}a - \cos k_{y}a$ for $B_{1g}$ symmetry and 
to $\sin k_{x}a\sin k_{y}a$ for $B_{2g}$ symmetry (where $a$ is the
lattice constant) \cite{Dev}. By fitting ARPES
data, the 
band-dispersion relation $\epsilon
(\vec{k})$ (in units of meV) for Nd$_{1.85}$Ce$_{0.15}$CuO$_{4-y}$  was found to be \cite{Mar}
\begin{eqnarray}
\epsilon (\vec{k}) = -460(\cos k_{x}a + \cos k_{y}a) + 220\cos k_{x}a\cos
k_{y}a \nonumber \\
  -70(\cos 2k_{x}a + \cos 2k_{y}a) 
  +60(\cos 2k_{x}a\cos k_{y}a \nonumber \\
  +\cos k_{x}a\cos 
  2k_{y}a)- 27
\end{eqnarray}

Figure 1 shows $B_{1g}$ and $B_{2g}$ Raman intensities at 8 K for
Nd$_{1.85}$Ce$_{0.15}$CuO$_{4-y}$ 
with $T_{c}$ = 22$\pm$1 K. The data are digitized from 
Ref.~\cite{Blu}. The solid lines 
are numerically calculated
curves using Eqs.~1 and 2, $\Gamma$ = 1.4 meV for the $B_{1g}$ channel, 
$\Gamma$ = 1.9 meV for the $B_{2g}$ channel, and  $\Delta$ = $3.59
(1-0.11\cos
4\theta)$ meV, where $\theta$ is measured from the Cu-O bonding
direction. It is remarkable that the calculated curves match very 
well with the experimental data. Furthermore, the minimum value $\Delta_{min}$ of this gap function 
is 3.2 meV, which is close to that (3 meV) deduced from the magnetic penetration 
depth \cite{Alff}. The optical reflectance of a similar
Nd$_{1.85}$Ce$_{0.15}$CuO$_{4-y}$
crystal with $T_{c}$ = 23 K  also shows a minimum gap of about 3.1 meV
at 10 K (Ref.~\cite{Homes97}). For comparison, we also numerically calculate Raman intensities 
in terms of a non-monotonic $d$-wave gap function: $\Delta$ = $3.25 (1.43\cos 2\theta-0.43\cos 6\theta)$ meV 
(dashed lines). This gap size, which is a factor of 1.7 larger than 
that extracted from ARPES \cite{Matsui}, matches the peak position of the $B_{2g}$ Raman
spectrum. It is apparent that the $B_{2g}$ Raman spectrum is inconsistent with 
any $d$-wave gap function with nodes along the Cu-Cu directions.

Further evidence for the nodeless $s$-wave
gap symmetry comes from the directional point-contact tunneling spectra of optimally 
doped Nd$_{1.85}$Ce$_{0.15}$CuO$_{4-y}$ ($T_{c}$ = 25 K).  
It was argued that single-particle tunneling experiments along 
the CuO$_{2}$ planes can probe the bulk 
electronic density of states since the mean free path is far larger than the 
thickness of the possibly degraded surface layer \cite{Pon}. Figure 2 shows the in-plane point-contact tunneling
spectra of Nd$_{1.85}$Ce$_{0.15}$CuO$_{4-y}$ measured along (100)
direction (Fig.~2a) and (110) direction (Fig.~2b), respectively.  The data are digitized 
from Ref.~\cite{Shan05}.  One can calculate point-contact tunneling
spectra using the Blonder-Tinkham-Klapwijk (BTK) theory \cite{BTK}. In this model, two
parameters are introduced to describe the effective potential
barrier ($Z$) and the superconducting energy gap $\Delta$. As a
supplement, the quasiparticle energy $E$ is replaced by $E-i\Gamma$,
where $\Gamma$ is the broadening parameter characterizing the finite
lifetime of the quasiparticles. Based on the BTK theory, Shan {\em et. al.}  
calculate the tunneling conductance in terms of the isotropic $s$-wave
gap function \cite{Shan05}. The agreement between the calculated curve and data is 
excellent for each tunneling spectrum. However, the gap sizes that
used to fit the tunneling spectra along the (100) and (110)
directions are slightly different. This implies that the gap is not
isotropic.

\begin{figure}[htb]
    \vspace{-0.3cm}
    \includegraphics[height=12cm]{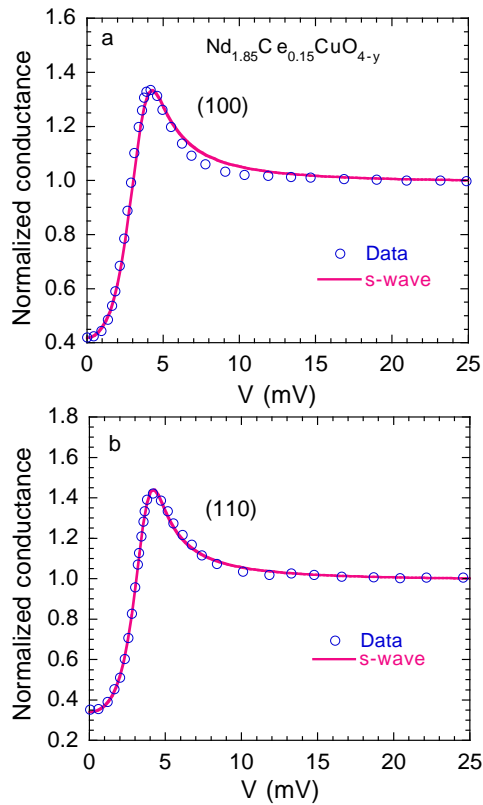}
     \vspace{-0.3cm}
 \caption[~]{In-plane point-contact tunneling
spectra of Nd$_{1.85}$Ce$_{0.15}$CuO$_{4-y}$ (
$T_{c}$ = 25 K) measured along (100)
direction (Fig.~2a) and (110) direction (Fig.~2b), respectively.  The data are digitized 
from Ref.~\cite{Shan05}.  The solid line in Fig.~2a is the
numerically calculated curve using $Z$ = 2.8, $\Gamma$ = 0.86 meV, and
$\alpha$ = 0 and the solid line in 
Fig.~2b is the calculated curve using $Z$ = 3.0, $\Gamma$ = 0.65 meV, and
$\alpha$ = $\pi$/4. The gap function used in the calculations is $\Delta$ = $3.52 (1-0.17\cos
4\theta)$ meV. }
\end{figure}

For the extended anisotropic
BTK model \cite{Tan}, another parameter $\alpha$ is introduced to
distinguish between different tunneling directions. In Fig.~2, we
compare the tunneling spectra with the calculated
curves based on the extended anisotropic
BTK model and an anisotropic $s$-wave gap function: $\Delta$ = $3.52 (1-0.17\cos
4\theta)$ meV. The solid line in 
Fig.~2a is the
numerically calculated curve using $Z$ = 2.8, $\Gamma$ = 0.86 meV, and
$\alpha$ =
0 and the solid line in 
Fig.~2b is the calculated curve using $Z$ = 3.0, $\Gamma$ = 0.65 meV, and
$\alpha$ =
$\pi$/4. It is apparent that the calculated curves are in
excellent agreement with the data. 

\begin{figure}[htb]
    \vspace{-0.5cm}
    \includegraphics[height=12cm]{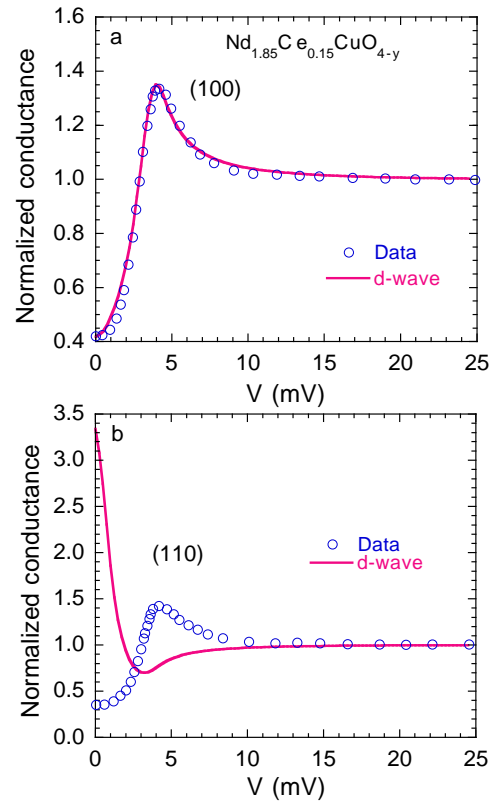}
     \vspace{-0.3cm}
 \caption[~]{Comparison of the tunneling spectra of NCCO  with the calculated
curves based on a nonmonotonic $d$-wave 
gap function: $\Delta$ = $3.0 (1.43\cos 2\theta-0.43\cos 6\theta)$ meV.   The solid
lines are the
numerically calculated curve using $Z$ = 3.0, $\Gamma$ = 0.65 meV, and
$\alpha$ = 0 for the spectrum along (100) direction and $\alpha$ = $\pi$/4 for the spectrum along
(110)
direction, respectively. }
\end{figure}

In Figure 3, we compare the tunneling spectra with the calculated
curves in terms of a nonmonotonic $d$-wave 
gap function: $\Delta$ = $3.0 (1.43\cos 2\theta-0.43\cos 6\theta)$ meV.
The solid
lines are the
numerically calculated curves using $Z$ = 3.0, $\Gamma$ = 0.65 meV,
$\alpha$ = 0 for the spectrum along (100) direction, and $\alpha$ = $\pi$/4 for 
the spectrum along (110) direction. One can see that, for the tunneling spectrum along (100) direction, the calculated curve coincides with
the data at high bias voltages but significant deviations occur at low 
bias voltages. For tunneling spectrum along (110) direction, ZBCP is clearly seen in the calculated curve, in
sharp contrast to the data. Therefore,  the tunneling spectra cannot
be explained by $d$-wave gap symmetry.

Finally, the most powerful way to distinguish between any $d$-wave and anisotropic
$s$-wave gap symmetries is to
study the response of a superconductor to nonmagnetic impurities or disorder. 
The nonmagnetic impurity pair-breaking effect is both bulk- and
phase-sensitive. 
This is because the rate of $T_{c}$ suppression by nonmagnetic impurities or defects
in a two-dimensional superconductor \cite{Op} is determined
by the value of the Fermi surface (FS) average $<\Delta
(\vec{k})>_{FS}$, which depends sensitively on the phase of the gap
function. More specifically, the rate is proportional to 
a parameter $\chi$ = $1-(<\Delta (\vec{k})>_{FS})^{2}$/$<\Delta^{2}(\vec{k})>_{FS}$.
It is clear that $\chi$ = 0 for isotropic $s$-wave superconductors
while $\chi$ = 1 for $d$-wave and $g$-wave superconductors. For the
anisotropic $s$-wave gap: $\Delta$ = $3.59 (1-0.11\cos
4\theta)$ meV, $\chi$
= 0.006. An equation to describe the pair-breaking effect by
nonmagnetic impurities (or defects) is given by \cite{Op}
\begin{equation}\label{pairbreak}
\ln \frac{T_{c0}}{T_{c}}= \chi[\Psi (\frac{1}{2} + \frac{0.122
          (\hbar\Omega_{p}^*)^{2}\rho_{r}}{T_{c}}) - \Psi
	  (\frac{1}{2})],
\end{equation}	  
where $\hbar\Omega_{p}^{*}$ is the renormalized plasma energy
\cite{Op,Rad} in units
of eV, $\rho_{r}$ is the residual resistivity in units of
$\mu\Omega$cm, and $\Psi$ is the digamma function. The renormalized
plasma energy can be independently determined from optical conductivity. 
Optical data of Pr$_{1.85}$Ce$_{0.15}$CuO$_{4-y}$ indicate $\hbar\Omega_{p}^{*}$ = 1.64
eV (Ref.~\cite{Homes}).  We will use this value of $\hbar\Omega_{p}^{*}$
to calculate $T_{c}$ as a function of the residual resistivity in
terms of both $d$-wave and an anisotropic $s$-wave gap function
inferred from the Raman spectra above.

\begin{figure}[htb]
    \vspace{-0.3cm}
    \includegraphics[height=6.5cm]{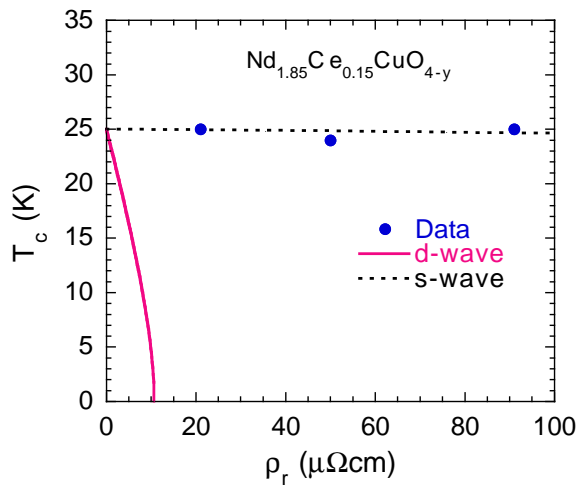}
     \vspace{-0.3cm}
 \caption[~]{$T_{c}$ as a function of residual resistivity in
optimally doped Nd$_{1.85}$Ce$_{0.15}$CuO$_{4-y}$. The data are from
Refs.~\cite{Alff,Shan05,On}. The solid line is numerically calculated curve in terms of any $d$-wave gap
and the dotted line is the calculated curve for  
an $s$-wave gap function proportional to $(1-0.11\cos
4\theta)$. }
\end{figure}

Figure 4 shows $T_{c}$ as a function of residual resistivity in
optimally doped Nd$_{1.85}$Ce$_{0.15}$CuO$_{4-y}$.
The data are from Refs.~\cite{Alff,Shan05,On}. 
The solid line is the numerically calculated curve in terms of any $d$-wave gap
and the dotted line is the calculated curve for  
an $s$-wave gap function proportional to $(1-0.11\cos
4\theta)$.  For the $d$-wave gap symmetry,
the parameter-free calculation (solid line) shows that $T_{c}$ will be suppressed
to zero at a very small residual resistivity of 10.7 $\mu\Omega$cm while the
measured $T_{c}$ is nearly independent of the residual resistivity.
For the $s$-wave gap proportional to $(1-0.11\cos
4\theta)$ or to $(1-0.17\cos
4\theta)$, the calculated $T_{c}$ is nearly independent
of the residual resistivity, in agreement with the data. Therefore, the data in
Fig.~4 rule out
any $d$-wave gap symmetry and 
unambiguously point towards nodeless $s$-wave gap symmetry.

In summary, the bulk-sensitive data of Raman scattering, 
optical conductivity, magnetic penetration depth, 
directional point-contact
tunneling spectra, and nonmagnetic pair-breaking effect in 
optimally electron-doped Nd$_{1.85}$Ce$_{0.15}$CuO$_{4-y}$ unambiguously
support
a nearly isotropic $s$-wave gap. The $s$-wave gap symmetry is
consistent with the earlier
\cite{Huang}
and recent \cite{Zhao09} conclusion that high-temperature
superconductivity in electron-doped cuprates is mainly caused by 
electron-phonon coupling.

$^{*}$Correspondence should be addressed to gzhao2@calstatela.edu


\begin{thebibliography}{99}
\bibliographystyle{prsty} 

\bibitem{Hardy}W. N. Hardy {\em et. al.},  Phys. Rev.
Lett. \textbf{70}, 3999 (1993).  

\bibitem{Jacobs}T. Jacobs {\em et. al.}, Phys. Rev. 
Lett. \textbf{75}, 4516 (1995).

\bibitem{Lee}S.-F. Lee {\em et. al.}, Phys. Rev. 
Lett. \textbf{77}, 735 (1996).

\bibitem{Chiao}M. Chiao {\em et. al.}, Phys. Rev. 
B \textbf{62}, 3554 (2000).

\bibitem{Bha}A. Bhattacharya {\em et. al.}, Phys.  Rev.  Lett.  \textbf{82}, 3132 (1999). 

\bibitem{Li}Q. Li {\em et. al.},  Phys.  Rev.  Lett.  \textbf{83}, 4160 (1999). This
phase-sensitive experiment is considered to be bulk-sensitive because 
the measured product of the critical current $I_{c}$ and the junction
resistance $R_{N}$ is in quantitative agreement with the
expected bulk value \cite{Zhao2001}.

\bibitem{Zhao2001} G. M. Zhao,   Phys.  Rev.  B {\bf 64}, 024503 (2001).

 \bibitem{Tsu} C. C. Tsuei, and J. R. Kirtley, Phys. Rev. Lett. {\bf 85}, 182
 (2000); A. D. Darminto {\em et. al.}, Phys. Rev. Lett. {\bf 94}, 167001 (2005). These phase-sensitive experiments are 
surface-sensitive because the inferred magnitude of the order
parameter from the measured critical current $I_{c}$ and junction
resistance $R_{N}$ is about two orders of magnitude too smaller than the
expected bulk value.  


\bibitem{Arm01} N. P. Armitage {\em et. al.}, Phys. Rev. Lett. {\bf 86}, 1126 (2001).

\bibitem{Matsui} H. Matsui {\em et. al.},  Phys.  Rev.  
Lett.  \textbf{95}, 017003 (2005).

\bibitem{Alff}  L. Alff {\em et. al.}, Phys.  Rev.  
Lett.  \textbf{83}, 2644 (1999). 

\bibitem{Pr} R. Prozorov, R. W.  Giannetta, P. Fournier,  and R. L. Greene,   Phys. Rev. Lett. {\bf 85},
03700 (2000).

 



 \bibitem{Kim}Mun-Seog Kim {\em et. al.},  Phys. Rev. Lett. {\bf 91}, 087001 (2002).

 
 

\bibitem{Kas}S. Kashiwaya {\em et. al.}, Phys. Rev. B {\bf 57}, 8680, (1998).


\bibitem{Bis}Amlan Biswas {\em et. al.}, Phys. Rev. Lett. {\bf 88},
207004 (2002).

\bibitem{Qaz} M. M. Qazilbash {\em et. al.}, Phys. Rev. B {\bf 68}, 024502 (2003).

\bibitem{Shan05}L. Shan {\em et. al.}, Phys. Rev. B {\bf 72}, 144506 (2005).

\bibitem{Shan08}L. Shan {\em et. al.}, Phys. Rev. B {\bf 77}, 014526 (2008).



\bibitem{Bet} J.  Betouras and R. Joynt,  Physica C {\bf 250}, 256 
(1995).

\bibitem{Mann}J. Mannhart and H. Hilgenkamp, Physica C 
\textbf{317-318}, 383 (1999).






 

\bibitem{Branch} D. Branch,  and  J. P. Carbotte, 
Phys. Rev. B. \textbf{52}, 603 (1995).

\bibitem{Dev} T. P. Devereaux, A. Virosztek,  and A. Zawadowski, Phys. Rev. 
B. \textbf{54}, 12523 (1996).

\bibitem{Mar}R. S. Markiewicz {\em et. al.},  Phys. Rev. B. \textbf{72}, 054519 (2005).

 \bibitem{Blu}  G. Blumberg {\em et. al.}, 
 Phys. Rev. Lett. \textbf{88}, 107002 (2002).
 
 \bibitem{Homes97} C. C. Homes, B. P. Clayman, J. L.  Peng, and  R. L. Greene, Phys.  Rev.  
 B  \textbf{56}, 5525 (1997). 



\bibitem{Pon}Ya. G. Ponomarev {\em et. al.}, Physica C {\bf} 243, 167 (1995). 

\bibitem{BTK} G. E. Blonder, M.  Tinkham,  and  T. M. Klapwijk,  Phys. Rev. B
{\bf 25}, 4515 (1982).

\bibitem{Tan} Y. Tanaka and  S. Kashiwaya,  Phys. Rev. Lett. {\bf 74},
3451 (1995).

\bibitem{Op} L. A. Openov,  Phys.  Rev.  B {\bf 58}, 9468 (1998).



\bibitem{Rad}R. J.  Radtke, K. Levin, H.-B. Schutter, M. R. Norman,  Phys.  Rev.  B {\bf 48}, 653
(1993).  


\bibitem{Homes} C. C. Homes {\em et. al.},  Phys.  Rev.  B  \textbf{74}, 214515 (2006). 



\bibitem{On}Y. Onose, Y. Taguchi, K. Ishizaka, and Y. Tokura, Phys. Rev. B {\bf 69}, 024504 
(2004). 

\bibitem{Huang} Q. Huang {\em et. al.}, Nature (London) {\bf 347}, 369
(1990).
\bibitem{Zhao09} G. M. Zhao, Phys. Rev. Lett. {\bf 103}, 236403 (2009).

\end{thebibliography}
\end{document}